\documentclass[12pt]{article}
\usepackage{epsf,amssymb,latexsym,enumerate,cite,shadow,array,color}
\usepackage{graphicx,wasysym}

\usepackage[pdftex]{hyperref}
\usepackage{url}

\newcommand{\be}{\begin{equation}}
\newcommand{\ee}{\end{equation}}

\newcommand{\rf}[1]{(\ref{#1})}
\newcommand{\vp}{\varphi}
\def\lesssim{\mathrel{\mathpalette\fun <}}
\def\gtrsim{\mathrel{\mathpalette\fun >}}
\def\fun#1#2{\lower3.6pt\vbox{\baselineskip0pt\lineskip.9pt
  \ialign{$\mathsurround=0pt#1\hfil##\hfil$\crcr#2\crcr\sim\crcr}}}
\relax
\begin{document}

\begin{titlepage}

\hskip 1cm

\vskip 3cm

\begin{center}
{\LARGE \textbf{On the problem of initial conditions\\ 
\vskip 10pt
for inflation}}\footnote{Based on an extended version of the invited talk at the conference ``Black Holes, Gravitational Waves and Spacetime Singularities,'' Specola Vaticana 9-12 May 2017}

\vskip 1.5cm

\

{\bf  Andrei Linde} 
\vskip 0.5cm
{\small\sl\noindent SITP and Department of Physics, Stanford University, \\
Stanford, CA
94305 USA\\
}
\end{center}
\vskip 1.5 cm

\

\begin{abstract}
 
I review the present status of the problem of initial conditions for inflation and describe several ways to solve this problem for many popular inflationary models, including the recent generation of the models with plateau potentials favored by cosmological observations.  
 \end{abstract}

\vspace{24pt}
\end{titlepage}

 \tableofcontents{}
\parskip 5.5pt

\parskip 5.5pt

\section{Initial conditions for inflation: A brief review} \label{review}

The theory of initial conditions in the early universe remains one of the most debated issues of modern cosmology. One can approach it at many different levels, e.g. in the context of quantum cosmology, or eternal inflation in string theory landscape, but it is always useful to go back to basics and check what is the status of this problem in the simplest models of inflation with one or two classical scalar fields. And even this may lead to misunderstandings since different people still have different ideas of what is inflation, and what is the meaning of the words ``initial conditions''. For example, for experts in numerical methods in GR,  initial conditions can be imposed at any time, whereas for people who want to understand the origin of the universe, initial conditions are related to the first moment when the classical description of the universe becomes possible. 

To clarify these issues,  let us remember what was the main problem with the hot Big Bang theory in this respect. In that model, the universe was born in the cosmological singularity, but it became possible to describe it in terms of classical space-time only when time was greater than the Planck time $t_{p} \sim 1$. At that epoch,  the temperature of matter was given by the Planck temperature $T_{p} \sim 1$, and the density of the universe was given by the Planck density $\rho_{p} \sim  1$. The size of the causally connected part of the universe at the Planck time was $ct_{p} \sim 1$. Each such part contained a single particle with the Planck temperature. The subsequent evolution was supposed to be nearly adiabatic, which means that the total number of particles in a comoving volume was approximately conserved. Thus this number from the very beginning was supposed to be greater than the total number of particles in the observable part of the universe, $n \sim 10^{{90}}$. This means that the universe at the Planck time consisted of $10^{90}$ causally disconnected parts. The probability that all of these independent parts emerged from singularity at the same time with the same energy density and pressure is smaller than $e^{-10^{90}}$, and it is even more complicated to explain why these parts would have the same density with an accuracy  better than $10^{-4}$.

Similar problem persisted in old \cite{Guth:1980zm} and new inflation \cite{Linde:1981mu}, where the universe was supposed to be large and hot from the very beginning, consisting of many causally disconnected parts. In this sense, these two models did not solve the problem of initial conditions. 

The situation changed dramatically with the invention of the chaotic inflation \cite{Linde:1983gd}. The main condition required in the simplest models of chaotic inflation was the existence of a single Planck size domain where the kinetic and gradient energy of the scalar field is few times smaller than its potential energy $V(\phi) \sim 1$. For sufficiently flat potentials, it leads to inflation, so the whole universe appears as a result of expansion of a single Planck size domain. According to  \cite{Linde:1983gd,Linde:1985ub,Linde:2005ht}, the probability of this process is not exponentially suppressed. After that, the universe described by the chaotic inflation scenario enters an eternal process of self-reproduction \cite{Linde:1986fd}.

This solution works for the simplest versions of the chaotic inflation scenario where inflation may start at the densities comparable with the Planck density. However, recent observational data \cite{Ade:2015lrj} favor inflationary models with plateau potentials, with the height of the plateau $V \sim 10^{{-10}}$. Such models include the GL model   \cite{Goncharov:1983mw}, in the Starobinsky model \cite{Starobinsky:1980te},  the Higgs inflation model  \cite{Salopek:1988qh,Bezrukov:2007ep}, and  the broad class of the cosmological attractor models \cite{Kallosh:2013hoa,Ferrara:2013rsa,Kallosh:2013yoa,Cecotti:2014ipa,Galante:2014ifa,Kallosh:2015zsa}, which generalize most of the previously proposed models with plateau potentials. Thus,  one could wonder  \cite{Ijjas:2013vea} whether it is possible to solve the problem of initial conditions for inflationary models of this type.

The answer to this question is two-fold. First of all, cosmological observations give us information only about the very last stages of inflation, and they tell us nothing about the beginning of inflation. I will describe several simple inflationary models where inflation begins at the Planck density with $V(\phi) \sim 1$ and ends by a slow-roll driven by a plateau potential with $V \sim 10^{{-10}}$. 

Moreover, in this paper I will show,  following \cite{Linde:2004nz,Linde:2014nna,Carrasco:2015rva,East:2015ggf,Kleban:2016sqm,Clough:2016ymm}, that one can easily solve the problem of initial conditions for  inflationary models  favored by the recent cosmological observations even if inflation is possible there only at $V \sim 10^{{-10}}$.  


\section{Models with a short plateau}

\begin{figure}[h!]
\begin{center}
\includegraphics[scale=0.55]{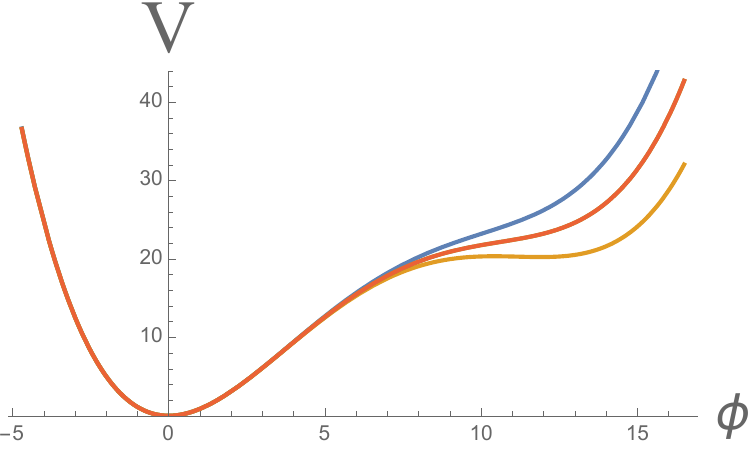}
\end{center}
\caption{\footnotesize The potential $V(\phi) = {m^{2}\phi^{2}\over 2}\,\bigl(1-a\phi +a^{2}b\,\phi^{2}\bigr)$ for $a = 0.12$ and $b = 0.30$ (upper curve), $b = 0.29$ (middle), and $b = 0.28$ (lower curve). The potential is shown in units of $m^{2}$, with $\phi$ in Planck units.  For  $b = 0.29$ (the middle curve), at the moment corresponding to $N=  58$ e-folding from the end of inflation one has $n_{s}= 0.965$ and $r = 0.012$, perfectly matching the Planck data. }
\label{chi}
\end{figure}

The simplest version of the chaotic inflation starting at $V \sim 1$ has the quadratic potential 
\be
{V}(\phi) = {m^{2}\phi^{2}\over 2} \ .
\ee
One can make a trivial modification of this model, by adding to it small terms proportional to $\phi^{3}$ and $\phi^{4}$.  
\be
{V}(\phi) = {m^{2}\phi^{2}\over 2}\,\Bigl(1-a\phi(1 +b\, a\, \phi )\Bigr) ,
\ee
For come values of parameters of this model, the potential acquires a short plateau,  see Fig. \ref{chi}. This helps to match the observational data, while still allows inflation to begin at $V = O(1)$ in Plank mass units, thus solving the problem of initial conditions for inflation.

Note that inflationary observations give us 3 main parameters: the amplitude of perturbations $A_{s}$, the spectral index $n_{s}$, and the tensor to scalar ratio $r$. By adjusting 3 parameters $m$, $a$ and $b$ in the modified potential one can easily describe the recent data describing the 3 parameters $A_{s}$,  $n_{s}$ and $r$. This is the simplest solution of the problem of initial conditions for inflation, compatible with the latest observational data. 

In what follows, we will show how one can go much further and solve the problem of initial conditions for a very broad class of models with plateau potentials, $\alpha$-attractors  \cite{Kallosh:2013yoa,Galante:2014ifa}. These models are very generic and economical; they can describe all presently available inflation-related observational data using a single parameter $m$ controlling the inflaton mass.

\section{$\alpha$-attractors}\label{basic}
There are many different ways to introduce the theory of $\alpha$-attractors, see  \cite{Kallosh:2013hoa,Ferrara:2013rsa,Kallosh:2013yoa,Cecotti:2014ipa,Galante:2014ifa,Kallosh:2015zsa}. On a purely phenomenological level, the main features of inflation in all of these models can be represented in terms of a single-field toy model with the Lagrangian  \cite{Galante:2014ifa,Kallosh:2015zsa}
 \be
 {1\over \sqrt{-g}} \mathcal{L} = { R\over 2}   -  {(\partial_{\mu} \phi)^2\over 2(1-{\phi^{2}\over 6\alpha})^{2}} - V(\phi)   \,  .
\label{cosmo}\ee
Here $\phi(x)$ is the scalar field, the inflaton.  The origin of the pole in the kinetic term can be explained in the context of hyperbolic geometry in supergravity and string theory.
The parameter  $\alpha$  can take any positive value. In the limit $\alpha \rightarrow \infty$ this model coincides with the standard chaotic inflation  with a canonically normalized field $\phi$ and the inflaton potential $V(\phi)$  \cite{Linde:1983gd}. 
However, for any finite values of $\alpha$, the field $\phi$ in \rf{cosmo} is not canonically normalized. It must satisfy the condition $\phi^2<6\alpha$, for the sign of the inflaton kinetic term to remain positive.  

We will assume that the potential $V(\phi)$ and its derivatives are non-singular for $\phi^2\leq6\alpha$. 
 Instead of the variable $\phi$, one can use a canonically normalized field $\vp$ by solving the equation ${\partial \phi\over 1-{\phi^{2}\over 6\alpha}} = \partial\vp$, which yields
\be\label{tanh} 
\phi = \sqrt {6 \alpha}\, \tanh{\varphi\over\sqrt {6 \alpha}} \ .
\ee
The full theory, in terms of the canonical variables, becomes
 \be
 {1\over \sqrt{-g}} \mathcal{L} = { R\over 2}   -  {(\partial_{\mu}\varphi)^{2} \over 2}  - V\big(\sqrt {6 \alpha}\, \tanh{\varphi\over\sqrt {6 \alpha}}\big)   \,  .
\label{cosmoqq}\ee
Note that in the limit $\phi \to 0$ the variables $\phi$ and $\varphi$ coincide; the main difference appears in the limit $\phi \to \sqrt{6 \alpha}$: In terms of the new variables, a tiny vicinity of the boundary of the moduli space at $\phi=\sqrt{6\alpha}$ stretches and extends to infinitely large $\varphi$. As a result, generic potentials $V(\phi) = V(\sqrt {6 \alpha}\, \tanh{\varphi\over\sqrt {6 \alpha}})$ at large $\vp$ approach an infinitely long dS inflationary plateau with the height corresponding to the value of $V(\phi)$ at the boundary:
\be
V_0 = V(\phi)|_{\phi = \pm \sqrt {6 \alpha}} \ .
\ee

To understand what is going on in this theory, let us consider, for definiteness, positive values of $\phi$ and study a small vicinity of the point $\phi = \sqrt {6 \alpha}$, which  becomes stretched to infinitely large values of the canonical field $\vp$  upon the change of variables $\phi \to \vp$. If the potential $V(\phi)$ is non-singular at the boundary  $\phi = \sqrt {6 \alpha}$, we can expand it in series with respect to the distance from the boundary:
\be
V(\phi) = V_{0} + (\phi-\sqrt {6 \alpha})\, V'_{0} +O(\phi-\sqrt {6 \alpha})^{2} \ .
\ee
where we denote $V'_{0} = \partial_{\phi}V |_{\phi = \sqrt {6 \alpha}}$. 

In the vicinity of the boundary $\phi=\sqrt {6 \alpha}$, the relation \rf{tanh} between the original field variable $\phi$ and the canonically normalized inflaton field $\vp$ is given by
\be\label{tanh2} 
\phi = \sqrt {6 \alpha}\, \left(1 - 2 e^{-\sqrt{2\over 3\alpha} \varphi }\right)\ ,
\ee
up to the higher order terms $O\bigl(e^{-2\sqrt{2\over 3\alpha} \varphi }\bigr) $. At $\vp \gg \sqrt \alpha$, these  terms are exponentially small as compared to the terms $\sim  e^{-\sqrt{2\over 3\alpha} \varphi }$, and the potential acquires the following asymptotic form:
\be\label{plateau}
V(\vp) = V_{0} - 2  \sqrt{6\alpha}\, V'_{0}\ e^{-\sqrt{2\over 3\alpha} \varphi } \ .
\ee
Note that the constant $2  \sqrt{6\alpha}\, V'_{0}$ in this expression can be absorbed into a redefinition (shift) of the field $\vp$. This implies that if inflation occurs at large $ \vp \gg \sqrt{\alpha}$, all inflationary predictions in this class of models of the potential $V(\phi)$ are determined only by the value of the potential $V_{0}$ at the boundary and the constant $\alpha$. For any values of $\alpha \lesssim 10$, the  amplitude of inflationary perturbations, the prediction for the spectral index $n_{s}$ and the tensor to scalar ratio $r$ match observational data under a single nearly model-independent condition 
\be
{V_{0}\over  \alpha} \sim m^{2} \sim 10^{{-10}} \ .
\ee
Thus the only parameter that is required to fit the present observational data is the parameter $m \sim 10^{{-5}}$ controlling the amplitude of the scalar perturbations of metric \cite{Kallosh:2015lwa}.

These results were explained in  \cite{Kallosh:2013hoa,Kallosh:2013yoa} and formulated in a particularly general way in \cite{Galante:2014ifa}:  The kinetic term in this class of models has a pole at the boundary of the moduli space. If inflation occurs in a vicinity of such a pole, and the potential near the pole has a finite and positive first derivative, all other details of the potential and of the kinetic term far away from the pole (from the boundary of the moduli space) become unimportant for making cosmological predictions. In particular, the spectral index depends solely on the order of the pole, and the tensor-to-scalar ratio relies on the residue  \cite{Galante:2014ifa}. All the rest is practically irrelevant, as long as the field after inflation falls into a stable minimum of the potential, with a tiny value of the vacuum energy, and stays there.
Stability of the inflationary predictions with respect to even very strong modifications of the shape of the potential outside a small vicinity of the boundary of the moduli space is the reason why these models are called cosmological attractors.

This new class of models accomplishes for inflationary theory something similar to what inflation does for cosmology. Inflation stretches the universe making it flat and homogeneous, and the structure of the observable part of the universe becomes very stable with respect to the choice of initial conditions in the early universe. Similarly, stretching of the moduli space near its boundary upon transition to canonical variables makes inflationary potentials very flat and results in predictions which are very stable with respect to the choice of the inflaton potential.

This addresses the often presented argument that the shape of the inflationary potential in large-field inflation and its cosmological predictions must be unstable with respect to higher order corrections to the potential at super-Planckian values of the field. In the new class of models, the range of the original field variables $\phi$ for $\alpha \lesssim 1$ is sub-Planckian, and the shape of the potential in terms of the canonical inflaton field is very stable with respect to the choice of the original potential \rf{plateau}, all the way to infinitely large values of the inflaton field. 

\begin{figure}[h!]
\begin{center}
\includegraphics[scale=0.55]{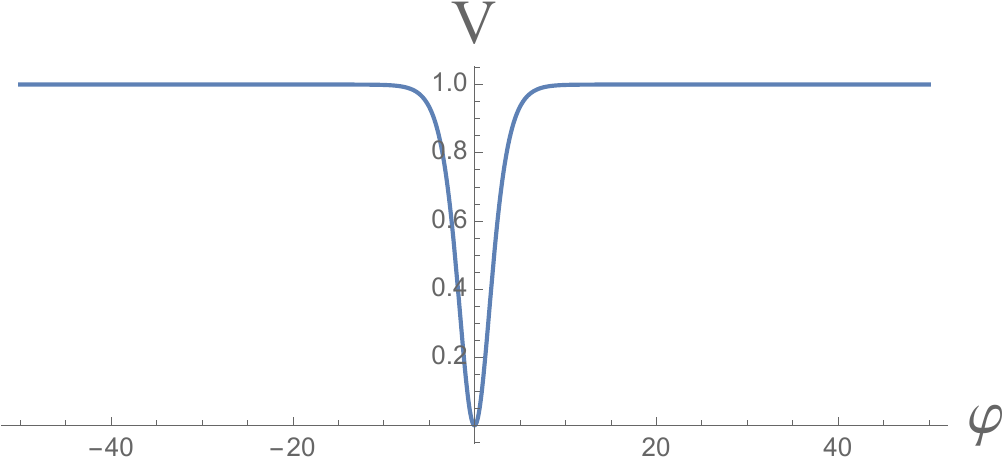}
\end{center}
\caption{\footnotesize The potential $V(\vp) =3\alpha  m^{2 }  \, \tanh^{2}{\varphi\over\sqrt {6 \alpha}}$ for $\alpha = 1$, shown in units of $3m^{2}$, with $\vp$ in Planck units.  For  $1/3< \alpha <10$   one has $n_{s}\sim 0.965$ and the tensor to scalar ratio $r$ is in the range from $3\times 10^{-2}$ to $10^{-3}$, providing good match to the Planck data. }
\label{F2}
\end{figure}

The simplest example of such theory is given by the model with $V(\phi) = m^{2}\phi^{2}$. In terms of the canonically normalized field $\vp$, the potential is given by
\be\label{shape}
 V(\varphi)=  3\alpha  m^{2 }  \, \tanh^{2}{\varphi\over\sqrt {6 \alpha}} . 
 \ee 
This is the simplest representative of the so-called T-models, with the T-shaped potential shown in  Fig. \ref{F2}

\section{$\alpha$-attractors with two fields}

\begin{figure}[h!]
\begin{center}
\includegraphics[scale=0.43]{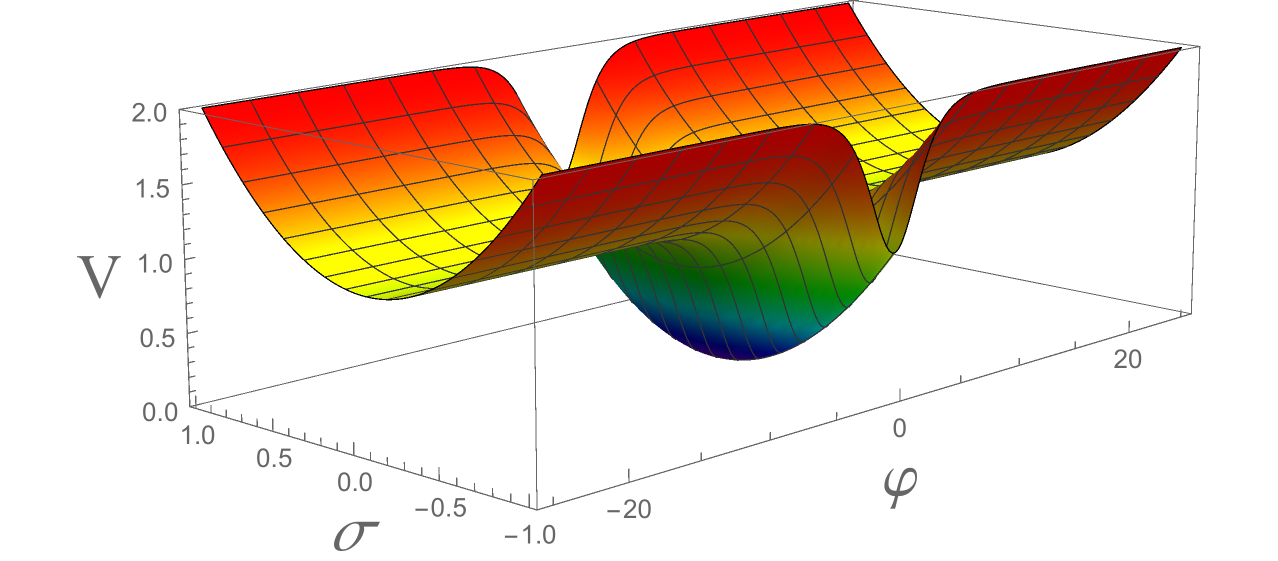}
\end{center}
\caption{\footnotesize The potential $V(\vp,\sigma)$ for $\alpha = 1$, shown in units of $3m^{2}$, with $\vp$ and $\sigma$ in Planck units. The shape of the potential along the valley $\sigma = 0$ is shown in Fig. \ref{F2}. }
\label{F3}
\end{figure}

Let us now consider an extended version of the $\alpha$-attractor model, adding to it a scalar field $\sigma$ with a canonically normalized kinetic term:
 \be
 {1\over \sqrt{-g}} \mathcal{L} = { R\over 2}   -  {(\partial_{\mu} \phi)^2\over 2(1-{\phi^{2}\over 6\alpha})^{2}} -  {(\partial_{\mu} \sigma)^2\over 2}  - {m^{2}\over 2} \phi^{2}   - {g^{2}\over 2}\phi^{2}\sigma^{2} - {M^{2}\over 2} \sigma^{2} .
\label{cosmo2}\ee
The inflaton potential becomes
 \be
 V(\varphi,\sigma) =    3\alpha (m^{2 }+g^{2}\sigma^{2})  \, \tanh^{2}{\varphi\over\sqrt {6 \alpha}} +{M^{2}\over 2} \sigma^{2} .
\label{cosmo3}\ee
Its shape is shown in Fig. \ref{F3}.

The potential depends on $|\vp|$. During inflation at  $|\varphi | \gg \sqrt\alpha$, one can use the asymptotic equation
\be
\tanh^{2}{|\varphi |\over\sqrt {6 \alpha}} = 1-4 e^{-\sqrt{2\over 3\alpha}|\varphi |} + O(e^{-2\sqrt{2\over 3\alpha}|\varphi |}) \ .
\ee
For notational simplicity, we will study positive values of $\vp$. The potential at  $\varphi \gg \sqrt\alpha$ is equal to 
\be
 V(\varphi,\sigma)=  3\alpha (m^{2 }+g^{2}\sigma^{2})  \, (1-4 e^{-\sqrt{2\over 3\alpha}\varphi})  +{M^{2}\over 2} \sigma^{2} , 
 \ee 
 up to exponentially small higher order terms $3\alpha (m^{2 }+g^{2}\sigma^{2}) O(e^{-2\sqrt{2\over 3\alpha}\varphi})$. The potential $ V(\varphi,\sigma)$ has a minimum with respect to $\sigma$ at $\sigma = 0$. The  inflaton potential at $\sigma = 0$ and  large $\varphi$ is  
 \be\label{ppotcan}
 V(\vp) = 3\alpha  m^{2 } \tanh^{2}{\varphi\over\sqrt {6 \alpha}} \ . 
  \ee
During inflation at $\vp \gg \alpha$, this potential with exponentially good accuracy coincides with the cosmological constant,
\be\label{ppot}
 V(\vp) \approx 3\alpha\, m^{2 } \ . 
 \ee
Mass squared of the canonically normalized field $\vp$ is given by the second derivative of $ 3\alpha  m^{2 } \tanh^{2}{\varphi\over\sqrt {6 \alpha}}$. At $\vp\ll \sqrt\alpha$  one has
\be
m^{2}_{\vp} = m^{2} \ ,
\ee
but at $\vp\gg \sqrt\alpha$ one finds 
$m^{2}_{\vp} = -8 m^{2} e^{-\sqrt{2\over 3\alpha}\varphi}$. Meanwhile
the mass of the field $\sigma$ at large $\vp$ approaches a constant value  
\be\label{ms}
m^{2}_{\sigma} = M^{2} + 6\alpha g^{2} \ .
\ee
and the potential asymptotically becomes a sum of the positive cosmological constant $3\alpha m^{2 }$, and a quadratic potential of the field $\sigma$:
\be\label{asympt}
 V(\varphi,\sigma)=  3\alpha m^{2 }+{M^{2} + 6\alpha g^{2}\over 2} \sigma^{2} . 
 \ee

The strength of interactions of  the  inflaton field with itself and with the field $\sigma$ during inflation at $\sigma = 0$ can be described in terms of the coupling constants of  canonically normalized fields, such as $\lambda_{\varphi,\varphi,\varphi,\varphi} = \partial^{4}_{\varphi}  V(\varphi, \sigma)_{{|}_{\sigma = 0}}$ or $\lambda_{\varphi,\varphi,\sigma,\sigma} = \partial^{2}_{\varphi} \partial^{2}_{\sigma} V(\varphi, \sigma)_{{|}_{\sigma = 0}}$. As one can easily see, all such couplings are suppressed by the exponentially small coefficient $e^{-\sqrt{2\over 3\alpha} \varphi }$. 

In other words, the inflaton field is ``asymptotically free''  \cite{Kallosh:2016gqp}. By that, we  mean the {\it exponentially small} strength of interactions of the field $\vp$ with all other fields  at large $\vp$, rather than the {\it logarithmically small} strength of interactions at large momenta, as in QCD.  Our conclusions apply to the models with any potential $V(\phi,\sigma)$ as long as this potential and its derivatives are non-singular at the boundary $\phi = \sqrt{6\alpha}$.
These unusual features of the new class of theories lead to stability of the plateau shape of the potential at large $\vp$ with respect to quantum corrections  \cite{Kallosh:2016gqp}.

\section{Solving the initial conditions problem for simplest single-field  $\alpha$-attractors}

Let us study the problem of initial conditions in the model with the $\alpha$-attractor  potential shown in Fig. \ref{F2}. As we can see, this potential is equal to the cosmological constant $\Lambda = 3\alpha m^{2} \sim 10^{{-10}}$ with exponentially good accuracy {\it everywhere along the infinitely long plateau from $-\infty$ to $+\infty$,} except for a narrow minimum at $|\vp| \lesssim \sqrt{6\alpha}$. This suggests that if we solve the problem of initial conditions for the exponential dS expansion of the universe containing normal matter and a positive cosmological constant, we will simultaneously solve the problem of initial conditions for inflation in the theories with plateau potentials.

This argument is nearly trivial, and one may wonder why it was not formulated in this simple form many years ago, because it turns the whole problem upside down: One may wonder whether there is any way {\it to escape} the exponential dS expansion of the universe containing normal matter and a positive cosmological constant? 

As an example, one may consider the present stage of the accelerated cosmological expansion. According to the simplest $\Lambda$CDM model, 70\% of matter in the universe is the positive vacuum energy, the cosmological constant. Because of that, the universe entered the stage of acceleration. There are many extremely large inhomogeneities in the universe, such as galaxies. The size of each of them does not grow at the same rate at the universe. Black holes do not grow as well. But neither galaxies nor the black holes can stop the general quasi-exponential expansion of the universe dominated by the cosmological constant. We have missed the point in time when this expansion could have been stopped. This could happen, for example, if at some stage we would find that we live in a locally overweight part of the universe, with the energy density of matter much greater than the cosmological constant. But this did not happen, and now it is too late: Density of matter decreases, while the cosmological constant does not, and the universe gradually enters the unstoppable dS expansion $a \sim e^{Ht}$.

The same can be expected for the early universe in the theory with the cosmological constant $V \sim 10^{-10}$, corresponding to the height of the inflationary plateau. In an expanding universe with any conventional equation of state, density of normal matter during the epoch of matter domination  drops down as $t^{-2}$. If the expansion continues longer than $t \sim V^{-1/2} \sim (\sqrt \alpha m)^{{-1}} \sim 10^{5}$ Planck times, i.e. longer than about $10^{-28}$ seconds, then the universe enters the stage of exponential expansion, after which the field $\vp$ slowly rolls down to its minimum while producing quantum fluctuations responsible for the structure formation in our universe.

Can the universe avoid inflation in such models? There are two obvious possibilities. First of all, for some reason the field $|\vp|$ may be born not on the infinite plateau but in the small vicinity off the minimum, with $|\vp| < \sqrt{6\alpha}$. Whereas it is possible, in the context of the models with long plateau potentials this possibility seems extremely unlikely  because of the infinitely large phase space for the canonical inflaton field with $|\vp| > \sqrt{6\alpha}$.

\begin{figure}[h!]
\begin{center}
\includegraphics[scale=0.55]{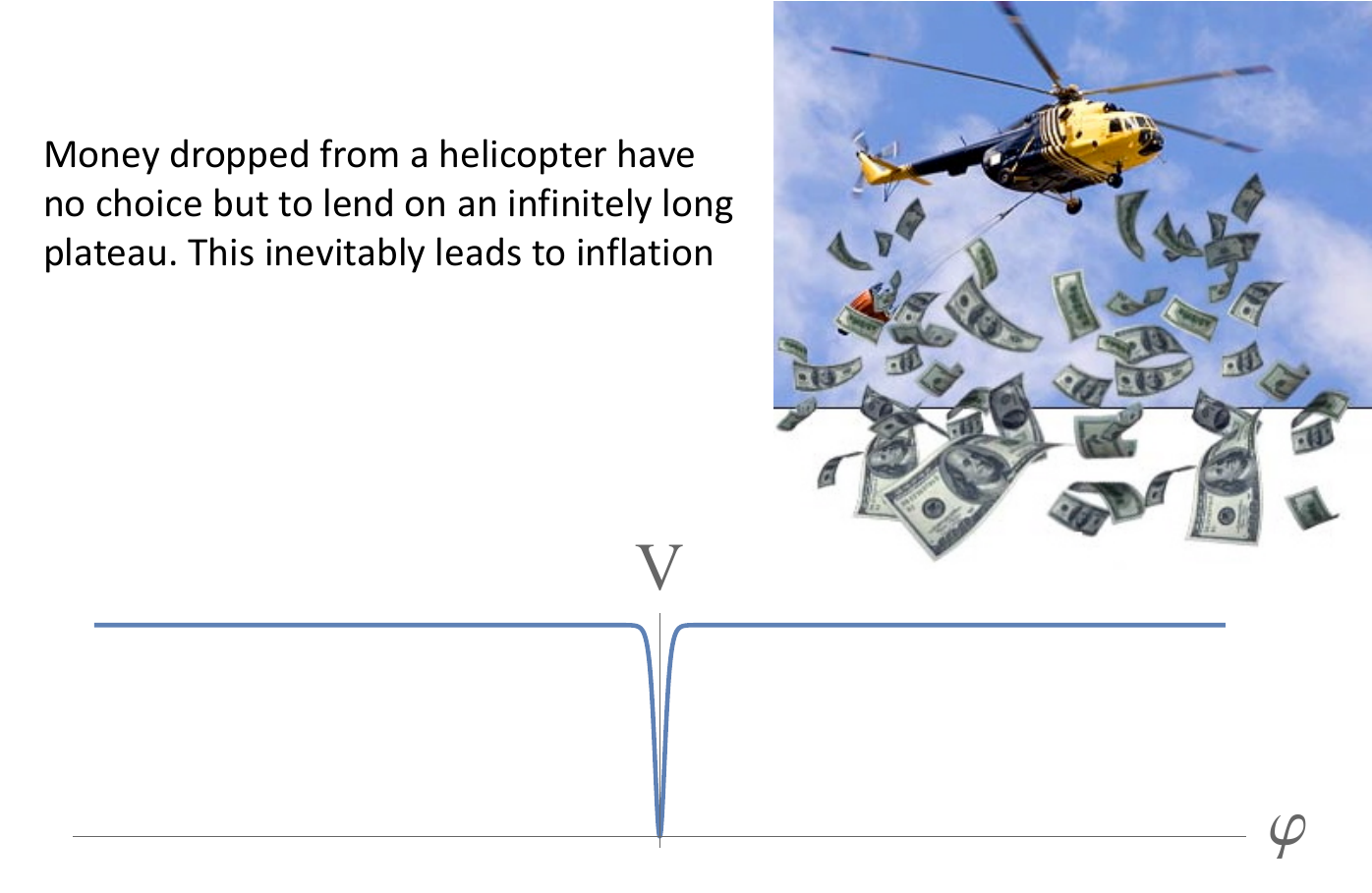}
\end{center}
\vskip -0.5cm 
\caption{\footnotesize Inflation in economy and in the universe. }
\label{F5}
\end{figure}

Following suggestion by Starobinsky, I would like to illustrate this argument made in  \cite{Carrasco:2015rva} by using an analogy with inflation in economy, see Fig. \ref{F5}. There is an often made statement that if the government drops lots of money from a helicopter, this may cause inflation. Of course, there is always a possibility that the money will be lost on its way. In our case, if we consider a theory with a plateau potential, and the scalar field $\phi$ drops in an expanding universe down from the state with the Planck energy density, then it is very difficult for it to end up in a narrow minimum of the potential and miss an infinite plateau. And if it falls to the plateau, then, just as in the situation with a positive cosmological constant background, it is difficult to avoid inflation.

The only other way to avoid inflation in this scenario is to assume that the whole universe collapses within $10^{-28}$ seconds. Indeed, if any part of an expanding universe continues expanding  longer than $10^{-28}$ seconds, the energy density of matter there becomes smaller than $V$, and it enters the stage of exponential expansion. {\it This means that all  universes, which are described by models with plateau potentials and can be explored by an observer with the lifespan greater than $10^{-28}$ seconds, should have passed the stage of inflation \cite{Carrasco:2015rva}.}  In other words, only virtual universes with the lifetime smaller than the inverse inflaton mass can avoid inflation.

This argument made in \cite{Carrasco:2015rva} was considerably advanced and strengthened by an analytical investigation and a computer simulation of a  universe expanding from the state with Planckian  density dominated by a grossly inhomogeneous scalar field $\vp$ \cite{East:2015ggf}, see also \cite{Kleban:2016sqm,Clough:2016ymm}. The simulations used sophisticated GR methods similar to the ones used for investigation of the black hole merger by LIGO. 

One of the features commonly used in numerical simulations is periodicity of the boundary conditions. This can be also interpreted as an investigation of a topologically non-trivial universe, like a torus. This is a somewhat unconventional approach, but it offers great advantages in studies of the problem of initial conditions for inflation, especially in the context of an open or flat universe. Indeed, one of the problems with the traditional approach to the Big Bang cosmology is that the open of flat universes were supposed to be infinitely large from the very beginning, with complicated correlations between infinitely many of their causally disconnected parts. No such problems appear if the universe is compact and topologically nontrivial \cite{Linde:2004nz,ZelStar,chaotmix,topol4,Coule:1999wg}. 

Consider, for example, an inhomogeneous expanding toroidal flat universe of the size of the horizon, which was of the same order as the Planck length $O(1)$ at the Planck time $O(1)$, when its total energy density was $O(1)$ in the Planck units. Since the potential energy density initially was $O(10^{{-10}}) \ll 1$, the universe was dominated by kinetic and gradient energy density of the scalar field, $\dot\vp^{2} \sim (\partial_{i}\phi)^{2} \sim 1$. In that case, the scale factor of the universe, which determined the size of the torus, was growing slower than the scale $H^{{-1}} \sim t$. And this means that the friction coefficient $\sim H$ rapidly became smaller than the original momenta of the scalar field inhomogeneities. In this situation, the effective equation of state of the inhomogeneities becomes similar to the equation of state of ultra-relativistic matter, which is not supporting the growth of inhomogeneities and black hole formation. Moreover, ultra-relativistic perturbations would travel many times around the small torus during the Hubble time, which would also support homogenization of the universe \cite{Linde:2004nz,chaotmix,topol4}. 

As a result, if the Planck size topologically non-trivial universe does not collapse as a whole within the Planck time, the chances that it will collapse later become small. The results of the computer simulations performed in  \cite{East:2015ggf} confirm this conjecture. We found that the original inhomogeneities may lead to a local process of  black hole formation shortly after the Planck time. But this process does  not affect large part of space, which continues expanding. This eventually results in a subsequent inflationary regime when the energy density of matter in an expanding part of the universe becomes smaller than the value of $V$ on the plateau.  This suggests a possible strengthening of our earlier formulation: most of the Planck size topologically nontrivial universes, which are described by the models with plateau potentials and have lifespan greater than the Planck time $\sim 10^{-33}$ seconds, eventually the stage of inflation.

Note that whereas the investigation of the problem of initial conditions was performed in  \cite{East:2015ggf} in the context of the models with plateau potentials, the actual computations evolved evolution in a rather limited range of the values of the inflaton field. Thus we expect that the final results should apply to many models of large field inflation.

\section{Singular $\alpha$-attractors and initial conditions for inflation}

\begin{figure}[h!]
\begin{center}
\includegraphics[scale=0.6]{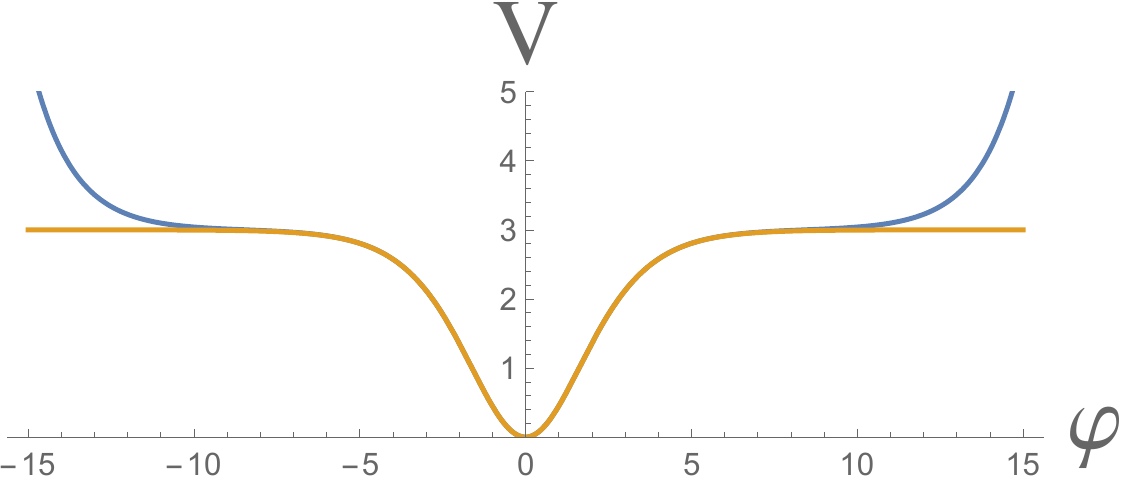}
\end{center}
\caption{\footnotesize The potential \rf{VVV} shown in units of $m^{2}$, for $\alpha = 1$ and for $A = 10^{-5}$ (upper curve) and $A = 0$ (lower curve). These two potentials nearly coincide at $\vp \approx 6$, corresponding to $N \sim 50$ e-foldings. }
\label{F6}
\end{figure}

Now let us return again to the basic single field $\alpha$ attractor models
 \be
 {1\over \sqrt{-g}} \mathcal{L} = { R\over 2}   -  {(\partial_{\mu} \phi)^2\over 2(1-{\phi^{2}\over 6\alpha})^{2}} - V(\phi)   \,  ,
\label{cosmoee}\ee
and
 \be
 {1\over \sqrt{-g}} \mathcal{L} = { R\over 2}   -  {(\partial_{\mu}\varphi)^{2} \over 2}  - V\big(\sqrt {6 \alpha}\, \tanh{\varphi\over\sqrt {6 \alpha}}\big)   \,  .
\label{cosmoqqee}\ee
and relax the usual assumption that the potential $V(\phi)$ is nonsingular at the boundary of the moduli space at $\phi = \pm \sqrt{6\alpha}$.  As an example, consider the potential
 \be
  V   = {m^{2}\phi^{2}\over 2}\left(1 +{A\over  1-{\phi^{2}\over 6\alpha}} \right)       =  3\alpha m^{2 }  \, \left(\tanh^{2}{\varphi\over\sqrt {6 \alpha}} +A \sinh^{2}{\varphi\over\sqrt {6 \alpha}}\right)  \ .
\label{VVV}\ee
For  $A \ll 1$, this potential has a long plateau, which ends at very large $\vp$, where the potential becomes exponentially steep, see Fig. \ref{F6}. 

In this theory, one can have a long stage of inflation dominated by the simplest $\alpha$-attractor potential $ 3\alpha m^{2 }   \,  \tanh^{2}{\varphi\over\sqrt {6 \alpha}}$. Indeed, as one can see from Fig. \ref{F6}, this potential  nearly coincides with the old  potential with $A = 0$ for  $\vp \approx 6$ corresponding to $N \sim 50$ e-foldings. But in the very early universe, at  asymptotically large values of  $\varphi$, the potential \rf{VVV} is given by
 \be 
  V  =  {A\, m^{2}\over 8}\  e^{\sqrt {2\over 3 \alpha}\,\vp}.
\label{VVV2}\ee
Cosmological evolution in such potentials is well known  \cite{Liddle:1988tb}: 
\be\label{power}
a(t) = a_{0}\ t^{3\,\alpha} \ .
\ee
 For $\alpha = 1/3$ we have a regime with $a \sim t$, which means that the energy of the homogeneous component of the scalar field decreased as $a^{{-2}}$, i.e. much more slowly than the energy of dust $\sim a^{{-3}}$ and of the relativistic gas $\sim a^{{-4}}$. This makes the solution of the problem of initial conditions proposed in \cite{Carrasco:2015rva,East:2015ggf} much simpler: If initially the energy of the homogeneous field was comparable to other types of energy, then in an expanding universe it gradually starts to dominate.  Meanwhile for  $\alpha > 1/3$ the power-law solution \rf{power} describes  inflation. It may begin already at the Planck density, which solves the problem of initial conditions in this class of models along the lines of \cite{Linde:1985ub}. 
 
 The scenario outlined above may have an additional non-trivial advantage. The amplitude of scalar perturbations $A_{s}$ in this scenario in the slow-roll approximation is given by 
 \be
 A_{s}  =  {V^{3}\over 12\pi^{2}V'^{2}}
 \ee
 In the theory without the singular term (i.e. with $A = 0$), the amplitude grows monotonically at large $\vp$ since $V$ approaches a plateau and $V'$ decreases. Meanwhile in the theory with the potential \rf{VVV}, the value of $V'$ exponentially decreases, and therefore $A_{s}(\vp)$ reaches some maximum and then falls down exponentially at large $\phi$. This effect suppresses the large wavelength amplitude of inflationary perturbations. With a proper choice of the potential, this may help to account for the observable decrease of the CMB anisotropy at small $l$ \cite{Contaldi:2003zv}.

\section{Solving the problem of initial conditions in two-field models}

\subsection{Initial conditions in models with several non-interacting scalars}
Two-field models can easily solve the problem of initial conditions not only for the large field inflation, but for small field inflation as well \cite{Linde:2014nna}. 
For example, let us start with a simple theory describing two non-interacting scar fields, $\phi$ and $\sigma$:
\be
U(\phi,\sigma) = V(\phi) + W(\chi) \ .
\ee
We will assume that the potential $W(\chi)$ contains a local minimum $\chi_{0}$ with very high energy density, and $W$ vanishes after the tunneling to the true vacuum. (More exactly, $W+V$ become vanishingly small, about $10^{-120}$, after the tunneling and the end of inflation, to account for the present tiny cosmological constant.) This is a standard feature of the string theory landscape scenario, which describes a huge number of such metastable states at all energy densities, many of which approach the Planck energy density \cite{Kachru:2003aw,Douglas:2003um,Susskind:2003kw}. Inflation in such metastable vacua is quite probable. Of course, this will be inflation of the old inflation type, so it does not solve any cosmological problems by itself, but it can create proper initial conditions for the slow-roll inflation, even if  $V(\phi)$ by itself can support slow roll inflation only for $V(\phi) \ll 1$.

Indeed, if the Hubble constant squared $H$ during the first stage of inflation supported  by the potential $W(\chi_{0})$ is greater than $V''$, fluctuations of the field $\phi$ are generated during this time, and after a while the universe becomes divided into exponentially large regions where the field $\phi$ takes all of its possible values. According to  \cite{Starobinsky:1986fx,Goncharov:1987ir,Linde:2005ht,Linde:1993xx,Linde:2006nw}, the probability to find a given value of the field $\phi$ at any given point in this regime is described by the probability distribution 
\be\label{distrib}
P(\phi) \sim \exp\left(-{24\pi^{2}\over W(\sigma_{0})}+{24\pi^{2}\over W(\chi_{0})+V(\phi)}\right) \approx \exp\left(-{24\pi^{2}V(\phi)\over W^{2}(\chi_{0})}\right)
\ee 
Therefore \rf{distrib} implies that at the moment prior to the decay of the metastable vacuum state in this model all values of the field $\phi$ such that $V(\phi)\ll 10^{{-3}} W^{2}(\chi_{0})$ will be equally probable. Note that in accordance with the Planck constraints on $r$, the value of the inflationary potential $V(\phi)$ during the last 60 e-foldings of inflation was smaller than $3\times 10^{-9}$. Therefore \rf{distrib} implies that initial conditions for the last 60 e-foldings of inflation in all theories $V(\phi)$ consistent with observations are quite  probable for $W^{2}(\chi_{0}) \gtrsim 10^{{-3}}$. Since the natural energy scale for the metastable vacua in the landscape can be almost as large as $O(1)$ in Planck units, one concludes that the initial conditions for the slow roll inflation in this scenario can naturally emerge after the stage of the false vacuum inflation in the landscape \cite{Linde:2005dd}.

One should add that the same scenario works even without any assumptions about the false vacuum inflation and landscape if one takes into account the possibility of the slow-roll eternal inflation in a theory of superheavy scalars $\chi$ and the light inflaton field $\phi$, see \cite{Linde:1987yb}.

\subsection{Initial conditions in models with several interacting scalars}

Consider now a theory with an effective potential 
\be
V(\phi,\chi) =  V(\phi) + W(\chi) +{g^2\over 2} \phi^2\chi^2
\ee
Here we assume that $V(\phi), W(\chi) \lesssim 10^{-9}$ are some potentials which cannot reach the Planckian values, which is the essence of the problem that we are trying to address. However, their interaction term ${g^2\over 2} \phi^2\chi^2$ can become $O(1)$, which would correspond to the Planck boundary. In that case the Planck boundary is defined by the condition
\begin{equation}\label{1}
{g^2\over 2} \phi^2\chi^2 \sim 1 \ ,
\end{equation}
which is represented by the set of four hyperbolas
\begin{equation}\label{2g}
 g |\phi ||\chi| \sim 1 \ .
\end{equation}
At larger values of $\phi$ and $\chi$ the density is greater than
the Planck density, so the standard classical description of
space-time is impossible there. In addition, the effective
masses of the fields should be smaller than $1$, and
consequently the curvature of the effective potential cannot be
greater than $1$. This leads to two additional conditions:
\begin{equation}\label{3}
  |\phi | \lesssim g^{-1}, ~~~~~~|\chi| \lesssim g^{-1}.
\end{equation}
We assume that $g \ll 1$. On the main part of the hyperbola \rf{2g} one either has $|\phi | \sim g^{-1} \gg |\chi \gg 1$, or $|\chi | \sim g^{-1} \gg |\phi \gg 1$. Consider for definiteness the first possibility. In this case, the effective mass of the field $\chi$, which is proportional to $g\phi$, is much greater than the effective mass of the field $\phi$, which is proportional to $g\chi$. Therefore in the beginning the field $\phi$ will move extremely slowly, whereas the field $\chi$ will move towards its small values much faster. Since its initial value on the Planck boundary is greater than $1$, the universe will experience a short stage of chaotic inflation determined by the potential ${g^2\over 2} \phi^2\chi^2$ with a nearly constant field $\phi$. After that, the first stage of inflation will be over, the field $\chi$ will oscillate, and within few oscillations it will give its energy away in the process of preheating \cite{KLS}. As a result, the classical field $\chi$ rapidly becomes equal to zero, the term ${g^2\over 2} \phi^2\chi^2$ disappears, and the potential energy reduces to $V(\phi)$. Soon after that, the second stage of inflation begins driven by the scalar field. As we already mentioned, it initial value can be $|\phi | \sim g^{-1} \gg |\chi \gg 1$. Thus the first stage of oscillations of the field $\phi$ provides good initial conditions for chaotic inflation driven by the scalar field $\phi$   \cite{Felder:1999pv}. Note that this effect is very general; it may occur for potentials $V(\phi)$ and $W(\chi)$ of {\it any} shape, either convex or concave, as long as they are small and at least one of them can support inflation. 

\subsection{Combining $\alpha$-attractors and simplest models of chaotic inflation} 

Now we will return to $\alpha$ attractors.  Consider, for example, the two-field $\alpha$-attractor model  \rf{cosmo2} with the potential shown in Fig. \ref{F3}. Solution of the problem of initial conditions in such models becomes trivial, being reduced to the theory of initial conditions in the simplest chaotic inflation models with a quadratic potential  \cite{Linde:1983gd,Linde:1985ub,Linde:2005ht}. Indeed, according to \rf{asympt}, the potential $V(\vp,\sigma)$ at   $\vp \gg \sqrt{6\alpha}$ reduces to  the quadratic potential with the cosmological constant,  $V(\varphi,\sigma)=  3\alpha m^{2 }+{M^{2} + 6\alpha g^{2}\over 2} \sigma^{2}$. Inflation in this model can start at the Planck density with $V(\varphi,\sigma) = O(1)$. Its early stages are driven by the field $\sigma$ with a quadratic potential, as in  \cite{Linde:1983gd,Linde:1985ub,Linde:2005ht}. This solves the problem of initial condition for the first stage of inflation driven by the field $\sigma$. When this stages completes, the field $\sigma$ vanishes, and the second stage of inflation driven by the field $\vp$ with the $\alpha$-attractor potential begins. It is this last stage that describes the latest 60 e-foldings of inflation, in agreement with the Planck data.

\section{Conclusions}
In this paper, I described several different ways to solve the problem of initial conditions for a broad class of inflationary models, including the theories with plateau potentials, where the last stage of inflation occurs when the potential can be many orders of magnitude below the Planck scale.  This shows, contrary to some recent claims in the literature \cite{Ijjas:2013vea},  that the inflationary models favored by the latest observational data, such as the GL model, the Starobinsky model, the Higgs inflation model, and the broad class of $\alpha$-attractors do not suffer from the problem of initial conditions. Moreover, under certain conditions this problem can be solved not only for large field models discussed in  \cite{East:2015ggf}, but for small field models as well.

I am very grateful to the organizers of the conference ``Black holes, Gravitational Waves and Spacetime Singularities'' for their hospitality.  This work  is supported by SITP and by the US National Science Foundation grant PHY-1720397.

\

\section{Appendix: Quantum creation of universes with non-trivial topology}

In the main body of the paper,  I was using simple intuitive arguments not requiring familiarity with the tools of quantum cosmology.  However, quantum cosmology \cite{DeWitt67} and the theory of quantum creation of the universe ``from nothing'' \cite{Vilenkin:1982de,Hartle:1983ai,Linde:1984ir,Vilenkin:1984wp} allows to look at the problem of initial conditions from a different perspective. 

Most of the related investigations described creation of a closed universe. But an initial size of a closed inflationary universe studied in  \cite{Vilenkin:1982de,Hartle:1983ai,Linde:1984ir,Vilenkin:1984wp} should be greater than $H^{-1}$, which is 5 orders of magnitude greater than the Planck length in the models considered above. In some cases, this may lead to exponential suppression of the probability of quantum creation of such universes, but this problem can be solved using anthropic considerations \cite{Vilenkin:1984wp}.
There is no exponential suppression of the probability of quantum creation of an open or flat compact  universe \cite{Linde:2004nz,ZelStar,Coule:1999wg}. I will briefly remind the corresponding results, following \cite{Linde:2004nz}.

Consider a flat compact universe having the topology of a
torus, $S_1^3$,
\begin{equation}\label{2}
ds^2 = dt^2 -a_i^2(t)\,dx_i^2
\end{equation}
with identification $x_i+1 = x_i$ for each of the three dimensions. We will assume
for  simplicity that $a_1 = a_2 = a_3 = a(t)$. In this case the curvature
of the universe and the Einstein equations written in terms of $a(t)$ will be
the same as in the infinite flat Friedmann universe with metric $ds^2 = dt^2
-a^2(t)\,d{\bf x^2}$. In our notation, the scale factor $a(t)$ is equal to the
size of the universe in Planck units $M_{p}^{{-1}} = 1$.

In order to derive the Wheeler-DeWitt equation  \cite{DeWitt67}  for the compact flat toroidal
universe, one should first consider the gravitational action 
\be S = \int dt\, d^3
x\,\sqrt{-g}\,\left(-{1\over 2}{R} +{1\over
2}\partial_{\mu}\phi\,\partial^{\mu}\phi-V(\phi)\right)\ee
and take into account that
the volume of the 3D box is equal to $a^{3}$.  Let us assume for a moment that $\phi$
is constant, which is the case if the field stays at the top of the potential, or at the dS plateau, as in the models which we discussed here. In this case one can represent the effective
Lagrangian for the scale factor as a function of $a$ and $\dot a$,
\begin{equation}\label{lagr}
L(a) = -3\dot a^{2} a - a^{3}V\ .
\end{equation}
Finding the corresponding Hamiltonian and using the Hamiltonian constraint $H
\Psi(a) =0$ yields the Wheeler-DeWitt equation
\begin{equation}\label{5}
\left[{d^2\over da^2} +12 a^4 V\right]\Psi(a) = 0
\end{equation}

For large $a$, the solution of Eq. (\ref{5}) can be easily obtained in the WKB
(semiclassical) approximation, $\Psi \sim a^{{-1}}\exp [\pm i {2a^3\sqrt
V\over\sqrt 3}]$; positive sign corresponds to an expanding universe. This
approximation breaks down at $a \lesssim V^{-1/6}$. At that time
the size of the universe is much greater than the Planck scale, but much
smaller than the Hubble scale $H^{-1}\sim V^{-1/2}$. The meaning of this
result, to be discussed below in a more detailed way, is that at $a \gg
V^{-1/6}$ the effective action corresponding to the expanding universe is very
large, and the universe can be described in terms of classical space and time.
Meanwhile at $a \lesssim V^{-1/6}$, the effective action becomes small, the
classical description breaks down, and quantum uncertainty becomes large. In
other words, contrary to the usual expectations, at $a \lesssim V^{-1/6}$ one
cannot describe the universe in terms of a classical space-time even though the
size of the universe at $a \sim V^{-1/6}$ is much greater than the Planck size,
and the density of matter as well as the curvature scalar in this regime
remains small, $R = 4 V \ll 1$.

\begin{figure}[h!]
\begin{center}
 \includegraphics[scale=0.55]{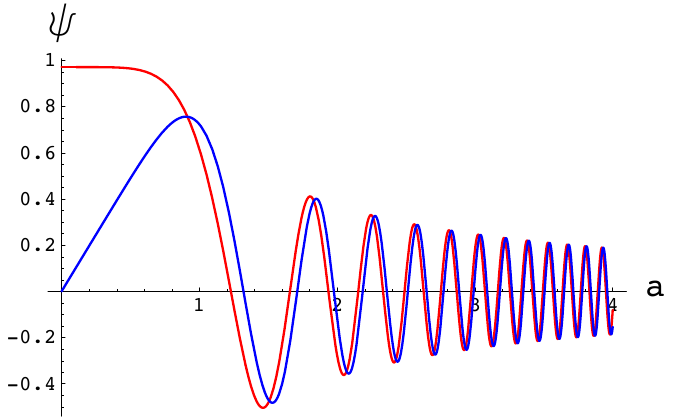}
\end{center}
\caption{\footnotesize Two eigenmodes of the Wheeler-DeWitt equation for the wave function of the flat compact toroidal universe. The scale factor $a$ is shown in units of $V^{-1/6}$.  }
\label{wdw}
\end{figure}

The general solution
for Eq. (\ref{5}) can be represented as a sum of two Bessel functions:
\begin{eqnarray}\label{funct} \Psi(a) =\beta\sqrt a \Bigl({\rm J}_{-{1\over
6}}\Bigl({2\sqrt V a^3\over \sqrt 3}\Bigr) + \gamma \, {\rm J}_{{1\over
6}}\Bigl({2\sqrt V a^3\over \sqrt 3}\Bigr)\Bigr),
 \end{eqnarray}
where $\beta$ and $\gamma$ are some complex constants, see Fig. \ref{wdw}.

Figure \ref{wdw} confirms that the WKB approximation is not valid at small $a$, and the ``cosmic clock'' starts ticking only at $a>V^{-1/6}$.

One can provide an alternative interpretation of this result, without invoking
the Wheeler-DeWitt equation.  By substituting the classical solution $a =
e^{Ht}$ into the effective Lagrangian (\ref{lagr}), one finds that the total action
of the universe is proportional to $\sqrt V a^{3}(t) \sim \sqrt V a^{3}(t) e^{{3Ht}}$

For $a < V^{-1/6}$, the action is smaller than 1, so the wave function does not oscillate. Not surprisingly, the total  energy of the universe at the critical time when $a$ becomes equal to $V^{-1/6}$ is of the same order as Hawking temperature $T_{H} =O(H)$, which corresponds to the typical energy of a single quantum fluctuation in dS universe. Thus the universe gradually emerges from nothing, and its wave function does not oscillate until its total energy reaches $O(H)$.

 Once the universe grows larger than $a \sim
V^{-1/6}$, its action rapidly becomes exponentially large, classical
description of the new-born universe becomes possible, and  its topology becomes irrelevant due to the magic of inflation.

Of course, if it is so easy to create a homogeneous universe, it may not be too difficult to create an inhomogeneous  universe as well. The most important conclusion of the investigation performed  in \cite{Linde:2004nz,ZelStar,Coule:1999wg} is that the probability of quantum creation of a compact homogeneous inflationary universe with non-trivial topology is not exponentially suppressed. Meanwhile the main result obtained recently in \cite{Carrasco:2015rva,East:2015ggf}  is that even if the new-born universe is grossly inhomogeneous, it typically becomes homogeneous at  later stages of the cosmological evolution.


\end{document}